\documentclass[twocolumn,superscriptaddress,showpacs,showkeys,amsmath,amssymb,prb]{revtex4}
\usepackage{graphicx}% Include Figure files
\usepackage{dcolumn}% Align table columns on decimal point
\usepackage{bm}% bold math
\usepackage{subfigure}

\usepackage{epsf}
\usepackage{epsfig}
\usepackage[latin1]{inputenc}   %?,?,?,... unter windows

\usepackage{amsmath}
\usepackage{amsthm}
\usepackage{amssymb}
\usepackage{epsfig}
\usepackage[bf,footnotesize]{caption}

\begin{document}

\title{Identifying vacancy complexes in compound semiconductors with positron annihilation spectroscopy: a case study of InN}
\author{Christian Rauch}
\email{christian.rauch@tkk.fi}
\affiliation{Department of Applied
Physics, Aalto University, P.O. Box 11100, FI-00076 Aalto, Espoo,
Finland}
\author{Ilja Makkonen}
\affiliation{Helsinki Institute of Physics and Department of Applied
Physics, Aalto University, P.O. Box 14100, FI-00076 Aalto, Espoo,
Finland}
\author{Filip Tuomisto}
\affiliation{Department of Applied Physics, Aalto University, P.O.
Box 11100, FI-00076 Aalto, Espoo, Finland}

\date{\today}

\begin{abstract}
We present a comprehensive study of vacancy and vacancy-impurity
complexes in InN combining positron annihilation spectroscopy and
ab-initio calculations. Positron densities and annihilation
characteristics of common vacancy-type defects are calculated using
density functional theory and the feasibility of their experimental
detection and distinction with positron annihilation methods is
discussed. The computational results are compared to positron
lifetime and conventional as well as coincidence Doppler broadening
measurements of several representative InN samples. The particular
dominant vacancy-type positron traps are identified and their
characteristic positron lifetimes, Doppler ratio curves and
lineshape parameters determined. We find that $V_{\text{In}}$ and
their complexes with $V_{\text{N}}$ or impurities act as efficient
positron traps, inducing distinct changes in the annihilation
parameters compared to the InN lattice. Neutral or positively
charged $V_{\text{N}}$ and pure $V_{\text{N}}$ complexes on the
other hand do not trap positrons. The predominantly introduced
positron trap in irradiated InN is identified as the isolated
$V_{\text{In}}$, while in as-grown InN layers $V_{\text{In}}$ do not
occur isolated but complexed with one or more $V_{\text{N}}$. The
number of $V_{\text{N}}$ per $V_{\text{In}}$ in these complexes is
found to increase from the near surface region towards the
layer-substrate interface.
\end{abstract}

% PACS, the Physics and Astronomy Classification Scheme.
%Use showkeys class option if keyword
                              %display desired
\maketitle

\section{Introduction}
InN is a significantly cation-anion mismatched semiconductor
compound~\cite{King2008} with many interesting properties and
promising applications in opto-, and high-speed
electronics~\cite{Bhuiyan2003}. Intrinsic point defects have been
accounted for multiple important mechanisms in the
material~\cite{Jones2007,Li2005,Piper2006,Tuomisto2007e,Reurings2010}.
Amongst them, In and N vacancies and their complexes are expected to
be the dominant intrinsic acceptors and donors, respectively,
according to latest density functional theory
calculations~\cite{Stampfl2000,Duan2008,Duan2009,Duan2009a,Walle2010}.
Nevertheless, unambiguous experimental evidence on their nature and
characteristics is still relatively scarce. This is due to
limitations stemming from intrinsic properties of InN, as well as
challenges related to the growth of high-quality material. Strong
surface electron accumulation in polar samples~\cite{Mahboob2004}
complicates the fabrication of Schottky contacts and therefore the
application of standard electrical characterization methods such as
deep level transient spectroscopy (DLTS) and capacitance voltage
(C-V) profiling. Additionally, the high conductivities common for
early as-grown InN layers together with the unavailability of bulk
material strongly limit the use of
electron paramagnetic resonance (EPR) based techniques.\\
Positron annihilation spectroscopy is a powerful method for the
investigation of vacancy type defects in
semiconductors~\cite{Saarinen1998} and largely not affected by the
above mentioned challenges. Positrons can get trapped and annihilate
at neutral and negatively charged open volume sites in the crystal
lattice due to a locally reduced Coulomb repulsion. This increases
the positron lifetime and narrows the momentum distribution of
annihilating electron-positron (e-p) pairs, both of which can be
measured by recording the emitted
annihilation $\gamma$ radiation.\\
While direct experimental evidence on the behavior of N vacancies
has been very limited, previous positron annihilation results show
that In vacancy ($V_{\text{In}}$) related defects are incorporated
in concentrations of $\sim$10$^{16}-10^{17}$cm$^{-3}$ during growth
of both molecular beam epitaxy
(MBE)~\cite{Oila2004,Laakso2004,Reurings2010b} and metal organic
chemical vapor deposition (MOCVD)~\cite{Pelli2006} InN. Although
$V_{\text{In}}$ related defect concentrations are found to be low in
the 10$^{16}$cm$^{-3}$ range in thick layers of high quality
as-grown material~\cite{Reurings2010b} the experimentally observed
concentrations are still by orders of magnitude higher than what
could be expected based on first-principles
calculations~\cite{Stampfl2000,Walle2010}. A strong influence of the
layer thickness on the vacancy concentrations has been
found~\cite{Oila2004}, together with a commonly observed increase
and qualitative change of the vacancy signal when approaching the
layer-substrate
interface~\cite{Oila2004,Pelli2006,Rauch2010,Reurings2010}. Growth
parameters such as polarity~\cite{Reurings2010} and
stoichiometry~\cite{Reurings2010b,Pelli2006} seem to have only minor
impact. This suggests that the vacancy formation in InN is not
dominated by thermal equilibrium processes but rather controlled by
mechanisms such as local strain, the vicinity to other point and
extended defects and/or limited surface diffusion during
growth~\cite{Reurings2010b}. Nevertheless, no direct correlation
between dislocation densities and vacancy formation has been found
so far~\cite{Reurings2010b,Wang2008}. In good agreement with the
behavior of negatively charged defects below the branch point
energy~\cite{King2008,Walukiewisz2001}, n-type doping of InN by
either Si~\cite{Rauch2010,Uedono2005} or high energy particle
irradiation~\cite{Tuomisto2007e} leads to an increasing
incorporation of $V_{\text{In}}$ related defects, while Mg-doped
samples (with lowered Fermi level positions) show only low
concentrations~\cite{Uedono2009}.\\
Unfortunately, the exact chemical identity of the vacancy defects
detected in previous studies has remained largely unknown. Recent
attempts however have shown that these limitations could in
principle be overcome by careful modeling of positron annihilation
parameters using ab-initio
calculations~\cite{Uedono2009,Hautakangas2006a,Makkonen2006a}.
Despite this, a comprehensive theoretical study of positron
annihilation in InN has still been missing.\\
In the following, we present an extensive identification of common
vacancy and vacancy-impurity complexes in InN combining positron
annihilation spectroscopy and ab-initio calculations. The employed
computational and experimental methods are presented in
section~\ref{section_Methods}. Calculated positron lifetimes and
momentum distributions of annihilating e-p pairs are shown in
section~\ref{Section_Theory} for a variety of vacancies and vacancy
complexes in InN. In section~\ref{Section_Experimental} these are
compared to experimental data from positron lifetime and Doppler
broadening measurements, and the dominant vacancy defects in
positron annihilation measurements of different as-grown, doped and
irradiated InN samples are identified. A critical discussion of the
results is presented in section~\ref{Section_Discussion} and
possible implications are outlined.

\section{Methods}\label{section_Methods}
\subsection{Experimental}
Positron annihilation measurements are performed using a
variable-energy (0.5-38~keV) slow positron beam. The Doppler
broadening of the e-p annihilation radiation is recorded with two
Ge-detectors with a combined Gaussian resolution function of
1.24~keV (0.66~a.u.) and 0.995~MeV (0.53~a.u.)
full-width-at-half-maximum (FWHM) at 0.511~MeV in the conventional
and coincidence Doppler setup~\cite{Lynn1977,Alatalo1996},
respectively. In the latter configuration, both annihilation photons
are detected simultaneously and only counted if energy conservation
is fulfilled (E$_{tot}$=1.022MeV). This significantly improves the
peak-to-background ratio up to $10^{6}$ and sharpens the detector
resolution. To assure statistical reliability, spectra of
$\sim$1$\times$$10^{6}$ and 3$\times$10$^{7}$ counts are accumulated
for each measurement point in the conventional and coincidence
setup, respectively.\\
In conventional Doppler mode the annihilation peak is analyzed using
the common integrated lineshape parameters~\cite{Saarinen1998} which
represent the annihilation fractions in the low (S) and high (W)
momentum parts of the spectrum. Integration windows of
$|p{_L}(S)|<$~0.4~a.u. ($\Delta$$E_{\gamma}<$~0.75~keV) and
1.5~a.u.~$<|p{_L}(W)|<$ 3.9~a.u. (2.9~keV~$<
\Delta$$E_{\gamma}<$~7.3~keV) are chosen for the S and W parameter,
respectively. In positron annihilation experiments the
time-integrated annihilation parameter $P_{exp}$ (e.g. average
positron lifetime, annihilation lineshape, S and W parameter)
constitutes a weighted sum of the characteristic values of the
present positron traps $P_{i}$ and the crystal lattice $P_{b}$,
\begin{equation}\label{DopplerFormular}
P_{exp}=\eta_{b}P_{b}+\Sigma_{i=1}^{n}\eta_{i}P_{i},
\end{equation}
with $\eta_{b}$ and $\eta_{i}$ being the positron annihilation
fractions of the lattice and the $i$-th defect, respectively.\\
The characteristic defect parameter $P_{i}$ of a certain positron
trap can hence be extrapolated from the recorded experimental signal
$P_{exp}$, if bulk parameter and the positron annihilation fractions
are known. The latter can be determined by decomposition of the
positron lifetime spectrum into its $n$ exponential decay
components~\cite{Saarinen1998}, or by a complementary measurement of
another time-integrated annihilation parameter whose set of
characteristic parameters is well known. The bulk parameter is
usually determined by measurement of a reference sample in which no
trapping to vacancy defects can be observed (confirmed by positron
lifetime measurements).

\subsection{Computational}\label{computational}
Our computational scheme~\cite{Makkonen2006a} is based on the
zero-positron-density limit of the two component density functional
theory~\cite{Boronski1986}. Valence electron densities are
calculated self-consistently using the local density approximation
(LDA) and projector augmented-wave method (PAW)~\cite{Bloechl1994}
implemented in the VASP code~\cite{Kresse1996}. All electronic
structure calculations are performed using a 96-atom InN wurtzite
supercell. Ionic positions are relaxed with a convergence criterium
of 0.01~eV/{\AA} for forces. Indium 4d electrons are treated as
valence and an energy cut-off of 400~eV is chosen. The Brillouin
zone is sampled with a 3$^{\text{3}}$ Monkhorst-Pack
\textbf{k}-point mesh.\\
After deriving the electron densities in the lattice, the positron
densities are solved independently in the calculated Coulomb
potential due to electrons and nuclei and the e-p correlation
potential. The calculation of the positron densities is performed in
the so-called "conventional scheme", i.e. $n_{+} \rightarrow 0$ is
used and the e-p correlation potential is approximated in the zero
positron density limit. This is only exact for the case of a
completely delocalized positron in the crystal lattice (bulk) but is
also justified for finite positron densities localized at a defect
site when considering the positron and its screening cloud of
electrons as a neutral quasi-particle which does not affect the
surrounding average electron density. The LDA~\cite{Boronski1986} in
the state-dependent scheme~\cite{Alatalo1996} is used for the
description of many-body effects in the calculation of positron
annihilation rates and the momentum distributions of annihilating
e-p pairs. In this approximation, the momentum density
$\rho(\textbf{p})$ of annihilating e-p pairs is expressed as
\begin{equation}
\rho(\textbf{p})=\pi r_{e}^{2}c \sum_{j}\gamma_{j}|\int
d\textbf{r}e^{-i\textbf{p}\cdot\textbf{r}}\psi_{+}(\textbf{r})\psi_{j}(\textbf{r})|^{2},
\end{equation}
where $r_{e}$ is the classical electron radius, $c$ the speed of
light and $\psi_{+}(\textbf{r})$, $\psi_{j}(\textbf{r})$ the
positron and electron wavefunctions, respectively. The summation is
performed over all occupied electron states and $\gamma_{j}$ is the
position independent electron-state-dependent enhancement
factor~\cite{Alatalo1996} which takes into account the increase in
annihilations due to the screening electron density around the
positron.\\
In order to be able to compare the calculated 3D momentum density to
1D experimental spectra we integrate the calculated spectra over the
wurtzite \textit{m}-plane. To account for the experimental detector
resolution the calculated momentum distributions are additionally
convoluted with a Gaussian resolution function of 0.53~a.u. and
0.66~a.u. FWHM for the comparison with coincidence and conventional
Doppler spectra, respectively. The line shape parameters S and W are
calculated from the spectra convoluted with 0.66~a.u. FWHM using the
above momentum windows. The positron lifetime $\tau$ is determined
as the inverse of the annihilation rate $\lambda$, which is given as
\begin{equation}
\lambda=\frac{1}{\tau}=\pi r_{e}^{2}c \int d\textbf{r}
n_{+}(\textbf{r})n_{-}(\textbf{r})\gamma(n_{-}(\textbf{r})),
\end{equation}
with $\gamma(n_{-}(\textbf{r}))$ the enhancement
factor~\cite{Alatalo1996} in the LDA.\\
In our calculations we made two further simplifications, i.e. all
defects are considered in the neutral charge state and forces due to
localized positrons are neglected. Adding negative (positive) charge
is expected to lead to an additional inwards (outwards)
relaxation~\cite{Duan2009a}, while the Coulomb force of localized
positrons is directed outwards~\cite{Makkonen2006a}. The exact
charge state of defect complexes is often unknown in experiments,
hence is not safe to be assumed without further prove. On the other
hand, only considering the positron-induced forces~\cite{Rauch2011}
leads to an overestimation of the outwards relaxation. Nevertheless,
careful testing showed that the effects of additional charge and
localized positrons on the annihilation characteristics can be
neglected compared to changes between different defects. Also,
trends in the calculated spectra are valid irrespective of which
approximation is used.

\section{Computational results}\label{Section_Theory}
The optimized lattice constants for wurtzite InN are calculated as
$a=3.510~\text{{\AA}}$, $c=1.610~\text{{\AA}}$ and $u=0.379$, in
good agreement with literature~\cite{IEE1994,Duan2009a}. The
positron density in the InN lattice is fully delocalized (see
Fig.~\ref{PositronDensity}(a)), and a positron lifetime of 157~ps is
determined.\\
The difference to the experimental lifetime value~\cite{Oila2004} of
$\sim$~180~ps stems from using the LDA enhancement factor. The
choice of the LDA is motivated by our focus on the calculation of
momentum distributions~\cite{Makkonen2006a}. In this approximation,
good agreement with experiments is achieved~\cite{Makkonen2006a} for
positron lifetime differences, $\Delta
\tau=\tau_{defect}-\tau_{bulk}$, rather than absolute values. When
analyzing computational positron lifetimes, it should be noted that
calculated values are sensitive to the size of the relaxed open
volume (i.e. overlap of positron and electron density), and small
differences in relaxation of a few
percent can lead to lifetime differences of a few ps.\\
The modeled momentum distribution of annihilating e-p pairs in the
InN lattice is displayed in Fig.~\ref{MomentumDis_undivided}. The
annihilation spectrum is symmetrical and is hence folded at its
center at 0~a.u (511~keV). In the low momentum part of the spectrum
annihilations with In~$5p$ and N~$2p$ valence electrons dominate
while contributions from tighter bound core electrons (In~$4d$ and
N~$2s$) become more important in the higher momentum part of the
spectrum.

\subsection{Isolated In vacancy}
\begin{figure}
\centering
\includegraphics[width=0.9\linewidth]{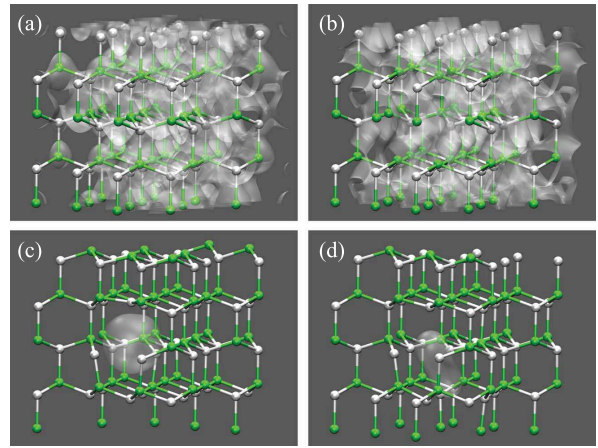}
  \caption{(Color online) Iso-surface plot of the calculated positron density (transparent sphere) in
  the relaxed lattice structure of the (a) InN lattice and (b)
  4$V_{\text{N}}$, (c) $V_{\text{In}}$, and (d) 2$V_{\text{In}}$ defects.
  Light grey (silver) and dark grey (green) balls indicate N
  and In ions, respectively.}
  \label{PositronDensity}
\end{figure}
\begin{figure}
\centering
\includegraphics[width=0.9\linewidth]{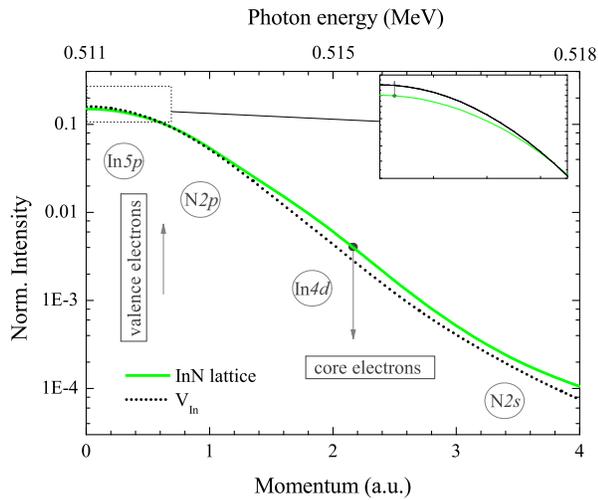}
  \caption{(Color online) Calculated momentum density of annihilating e-p pairs in the delocalized state of the InN lattice and trapped at the $V_{\text{In}}$, respectively. Trends in the spectra are indicated with respect to the InN lattice.}
  \label{MomentumDis_undivided}
\end{figure}
\begin{figure} \centering
\includegraphics[width=1\linewidth]{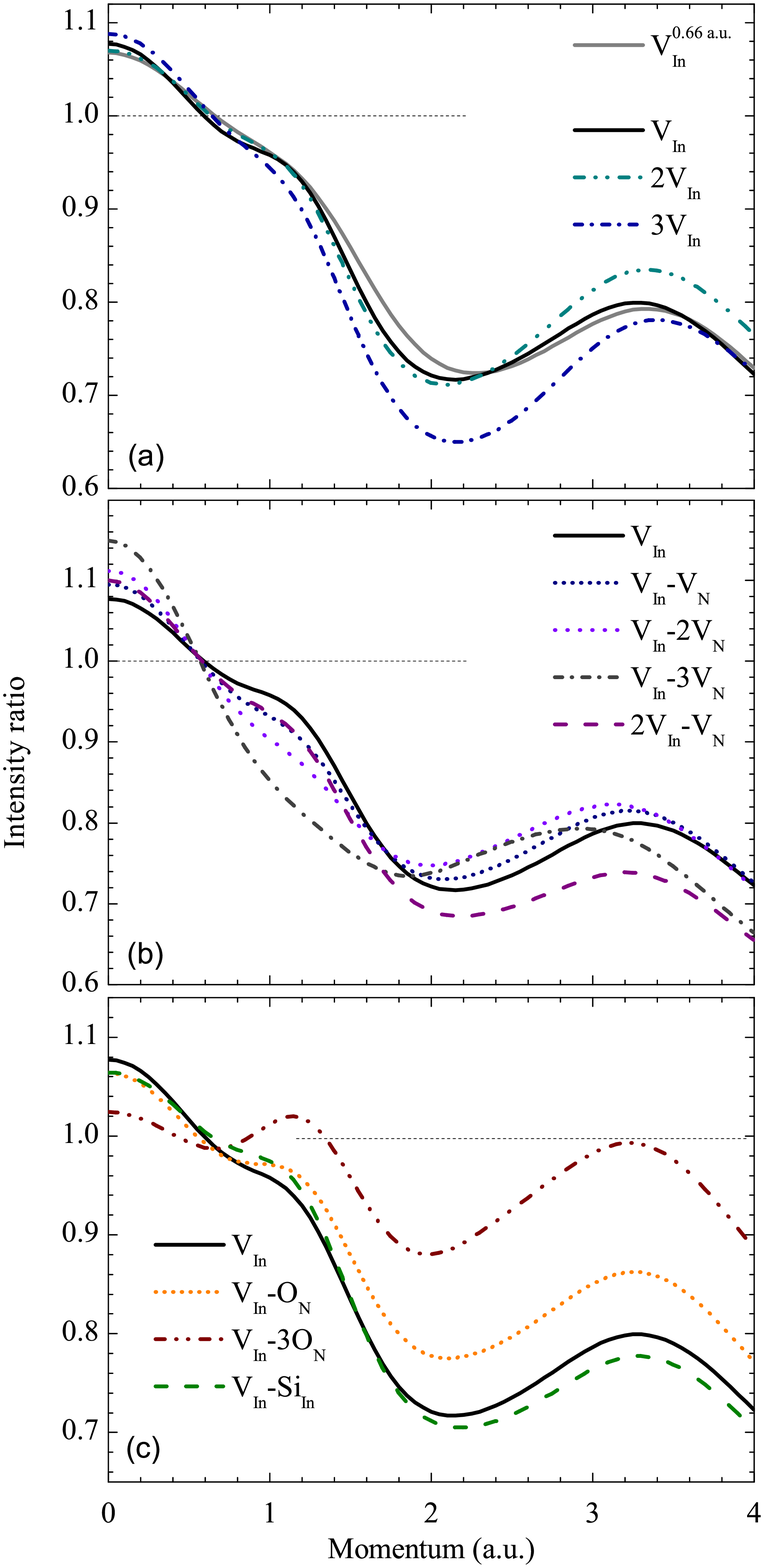}
  \caption{(Color online) Ratio curves of the calculated momentum densities of annihilating e-p pairs
  in selected vacancy complexes in InN. All spectra are convoluted with a Gaussian of 0.53~a.u. FWHM (except $V_{\text{In}}^{0.66~a.u.}$, FWHM~=~0.66~a.u.) and divided by the
  momentum density spectrum of the InN lattice.}
  \label{THEORY_Combined}
\end{figure}
\begin{figure}
\centering
\includegraphics[width=0.9\linewidth]{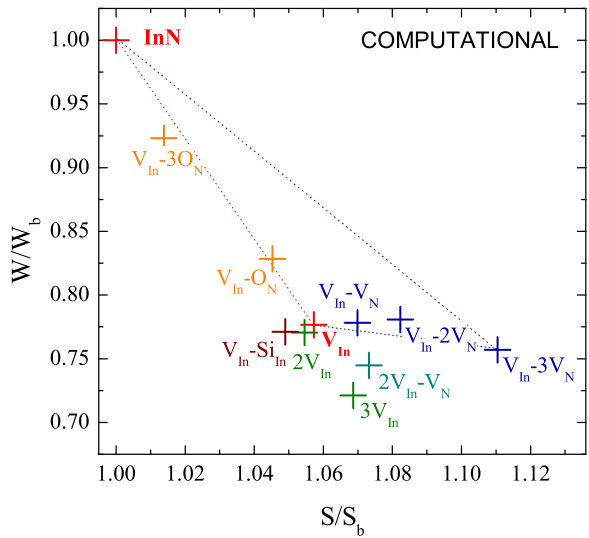}
  \caption{(Color online) SW-plot of conventional lineshape parameters S and W for the calculated momentum distributions of annihilating e-p pairs trapped at different vacancy type defects in InN.}
  \label{SW_THEORY}
\end{figure}

The relaxed defect geometry of the $V_{\text{In}}$ is shown in
Fig.~\ref{PositronDensity}(c) together with the calculated positron
density in the lattice. An outwards relaxation of the neighboring N
ions from their initial positions is observed in good agreement with
Duan \textit{et al.}~\cite{Duan2009a}. The positron density shows a
strong localization at the defect site which means that
$V_{\text{In}}$ act as efficient positron traps in InN. The
calculated positron lifetime difference to the InN lattice is 85~ps,
which is in good agreement with the $\sim$~80~ps observed in
experiments~\cite{Rauch2010,Oila2004}. Results obtained with the
atomic superposition (ATSUP) method~\cite{Puska1983} show for
positrons trapped at the $V_{\text{In}}$ increased annihilations
with loosely bound valence electrons (In~$5p$ and N~$2p$) while
annihilations with tighter bound core electrons (most importantly
In~$4d$) decrease significantly. This leads to an overall narrowing
of the annihilation peak compared to the InN lattice (see
Fig.~\ref{MomentumDis_undivided}).\\
For a more detailed analysis of the momentum spectrum we examine its
so-called ratio curve in Fig.~\ref{THEORY_Combined}(a), in which the
calculated momentum distribution is displayed divided by the
spectrum for the defect-free InN lattice to accentuate the
defect-induced changes.
The ratio curve for the $V_{\text{In}}$ exhibits a distinct line
shape with a maximum of roughly 1.08 at the peak center region (0
a.u.). For momenta above 0.6~a.u. the spectrum drops below 1 and an
articulate shoulder is visible at 1.2~a.u. which has been determined
by ATSUP calculations to stem from from annihilations with N~$2p$
electrons. At around 3.3~a.u. a second broad peak appears with an
intensity of around 0.8 relative to the InN
lattice.\\
The characteristic relative line shape parameters (for the
conventional detector resolution of 0.66~a.u.) of the
$V_{\text{In}}$ momentum distribution spectrum are calculated as
1.057 and 0.78 for S and W, respectively. Fig.~\ref{SW_THEORY} shows
the calculated values plotted in a SW-plot.

\subsection{In vacancy complexes}
Recent DFT results~\cite{Duan2009a} suggest that $V_{\text{In}}$
would form complexes with each other and predict a significant
decrease in formation energies when going from a single
$V_{\text{In}}$ to 2$V_{\text{In}}$ and 3$V_{\text{In}}$ complexes.
Motivated by these results we calculate the momentum densities and
positron lifetimes of the 2$V_{\text{In}}$ and 3$V_{\text{In}}$ in
their most favorable configurations~\cite{Duan2009a}. In the
2$V_{\text{In}}$ pair the two vacancies are located out-of-plane
(with respect to the c-plane) on next-nearest neighbor positions,
and for the 3$V_{\text{In}}$ complex an additional vacancy is added
in the c-plane sharing the same N atom. As expected, both
2$V_{\text{In}}$ and 3$V_{\text{In}}$ complexes show localized
positron densities. Although the $V_{\text{In}}$ are only situated
on next-nearest neighbor sites their open volumes are connected,
showing e.g. a handlebar-like structure for the positron density at
the 2$V_{\text{In}}$ pair (see Fig.~\ref{PositronDensity}(d)).\\
Positron annihilation characteristics of the 2$V_{\text{In}}$ defect
are very similar to the isolated $V_{\text{In}}$. The calculated
positron lifetime is equal and the lineshape parameters are very
close to the isolated case, with $\Delta \tau$~=~85~ps, S~=~1.055
and W~=~0.77 (see Fig.~\ref{SW_THEORY}). In the ratio curve in
Fig.~\ref{THEORY_Combined}(a), a slight increase of the peak at
3.3~a.u. is visible but the absolute intensity at these momentum
values is already reduced by $\sim$~10$^{3}$ compared to the peak
maximum (see Fig.~\ref{MomentumDis_undivided}) and discrimination
(in experiments) is hence considerably complicated.\\
For the 3$V_{\text{In}}$ complex we calculate an increase in the
positron lifetime of roughly 10~ps compared to the isolated
$V_{\text{In}}$ ($\Delta \tau$~=~94~ps). Additionally, the ratio
curve changes significantly with an increased peak maximum and a
more pronounced drop at 2.1~a.u. This is also visible in the change
of the lineshape parameters to S~=~1.069 and W~=~0.72. Nevertheless,
further analysis shows that the relative lineshape of the
3$V_{\text{In}}$ and $V_{\text{In}}$ are very similar and the
spectrum of the $V_{\text{In}}$ can nearly be reproduced from the
3$V_{\text{In}}$ spectrum by assuming a positron annihilation
fraction of $\eta$~$\approx$~0.8 (see Eq.~\ref{DopplerFormular}).
This is also visible in the SW-plot (Fig.~\ref{SW_THEORY}) in which
the characteristic point of the 3$V_{\text{In}}$ falls on an
extension of the line connecting the InN lattice point and the
$V_{\text{In}}$. For measurement points lying between the
$V_{\text{In}}$ and the InN lattice, the 3$V_{\text{In}}$ and
$V_{\text{In}}$ can therefore only be distinguished with precise
knowledge of the positron annihilation fraction. This is in practice
limited by the achievable accuracy in the separation of lifetime
components in positron lifetime measurements.

\subsection{N vacancy complexes}
For most of the Fermi level regions, $V_{\text{N}}$ in InN are
supposed to be donors in the 3+ charge state~\cite{Stampfl2000}
(hence repelling positrons) and possess a substantially smaller open
volume compared to the $V_{\text{In}}$, due to the large size
difference between In and N atoms. Both suggests strongly that
isolated $V_{\text{N}}$ can not act as positron traps in InN.
Nevertheless, this may not count for larger complexes of several
$V_{\text{N}}$ which possess a positive binding energy (compared to
isolated $V_{\text{N}}$) according to recent first-principles
calculations~\cite{Duan2008}. The increased open volume in these
defects could in principle promote a localization of the positron
density. Additionally, the $V_{\text{N}}$ complexes are assumed to
adapt negative charge states for elevated Fermi level positions
which additionally supports trapping.\\
To assess whether $V_{\text{N}}$ or its complexes could act as
positron traps in InN, we calculated the electron and positron
densities for the isolated $V_{\text{N}}$ and $V_{\text{N}}$
complexes of up to 4 N vacancies in the neutral charge state. The
defect structure and positron density for the 4$V_{\text{N}}$
complex are shown in Fig.~\ref{PositronDensity}(b), exemplarily. We
find that the positron density for all calculated $V_{\text{N}}$
complexes is clearly delocalized and no bound positron state exists
at $V_{\text{N}}$ or $V_{\text{N}}$ complexes. Hence, isolated
$V_{\text{N}}$ and pure $V_{\text{N}}$ clusters (in the neutral
charge state) cannot be detected using positron annihilation
spectroscopy. For the case of negatively charged $V_{\text{N}}$
complexes the formation of weakly localized hydrogenic positron
states around the defects is expected which possess very similar
annihilation characteristics to the defect free lattice but might be
detectable through their specific temperature
behavior~\cite{Saarinen1998}.

\subsection{Mixed In and N vacancy complexes}
Apart from pure $V_{\text{In}}$ and $V_{\text{N}}$ complexes, a
positive binding energy is predicted between isolated
$V_{\text{In}}$ and $V_{\text{N}}$, which should promote the
formation of divacancies ($V_{\text{In}}$-$V_{\text{N}}$) and larger
mixed vacancy complexes in the material~\cite{Duan2009a}. We
calculate the positron densities and positron annihilation
characteristics for the relaxed lattice structures of a variety of
mixed vacancy complexes in InN, namely
$V_{\text{In}}$-$V_{\text{N}}$, $V_{\text{In}}$-2$V_{\text{N}}$,
2$V_{\text{In}}$-$V_{\text{N}}$ and $V_{\text{In}}$-3$V_{\text{N}}$.
As expected, the increased open volume leads in all structures to a
clear localization of the positron density at the defect site. The
calculated positron lifetimes are close to the $V_{\text{In}}$
lifetime for the $V_{\text{In}}$-$V_{\text{N}}$ and
$V_{\text{In}}$-2$V_{\text{N}}$, with $\Delta \tau$~=~86 and 88~ps,
but increase to 99 and 109~ps for the
$V_{\text{In}}$-3$V_{\text{N}}$ and 2$V_{\text{In}}$-$V_{\text{N}}$,
respectively. Fig.~\ref{THEORY_Combined}(b) shows the ratio curves
of the computed momentum distributions. A systematic trend compared
to the isolated $V_{\text{In}}$ is visible in the spectra when
adding an increasing amount of $V_{\text{N}}$ around a single
$V_{\text{In}}$. A strong increase of the zero momentum maximum to
over 1.15 for the $V_{\text{In}}$-3$V_{\text{N}}$ is visible which
is related to the increase in open volume. At the same time, the
intensity of the shoulder at 1.2~a.u. decreases with increasing
number of $V_{\text{N}}$ until it entirely disappears for the
$V_{\text{In}}$-3$V_{\text{N}}$. The ratio curve of the
2$V_{\text{In}}$-$V_{\text{N}}$ is close to the
$V_{\text{In}}$-$V_{\text{N}}$ for lower momentum values but starts
to deviate at around 1.4~a.u. with lower intensities at higher
momenta, due to reduced annihilation with
In~$4d$ electrons.\\
The characteristic line shape parameters of the calculated complexes
develop accordingly and show a clear deviation from the
InN~-~$V_{\text{In}}$ line in Fig.~\ref{SW_THEORY}, which is already
visible for the divacancy and intensifies with increasing number of
$V_{\text{N}}$. S and W values for the
$V_{\text{In}}$-$V_{\text{N}}$, 2$V_{\text{In}}$-$V_{\text{N}}$,
$V_{\text{In}}$-2$V_{\text{N}}$ and $V_{\text{In}}$-3$V_{\text{N}}$
are determined as 1.070 and 0.78, 1.082 and 0.78, 1.073 and 0.75,
and 1.115 and 0.76, respectively.

\subsection{Vacancy-impurity complexes}
III-vacancy donor-impurity complexes are well established in
III-Nitrides. In GaN, a binding energy of 1.6~eV has been
determined~\cite{Tuomisto2006} between $O_{\text{N}}$ and
$V_{\text{Ga}}$. The binding energy between between $V_{\text{Ga}}$
and $Si_{\text{Ga}}$ is considerably smaller ($V_{\text{Ga}}$ and
$Si_{\text{Ga}}$ are only next-nearest neighours) but still
positive~\cite{Neugebauer1996}. Recent ab-initio
calculations~\cite{Duan2009} predict also in InN a reduction of the
defect formation energy for $V_{\text{In}}$-$O_{\text{N}}$ complexes
compared to the isolated case, and
even stronger for the case of $V_{\text{In}}$-3$O_{\text{N}}$.\\
Hence, we take a look at the relaxed defect structures and positron
densities for the $V_{\text{In}}$-$O_{\text{N}}$,
$V_{\text{In}}$-3$O_{\text{N}}$ and
$V_{\text{In}}$-$Si_{\text{In}}$. For all complexes the positron
density is localized at the defect site. The calculated positron
lifetimes to the InN lattice are  again very close to the isolated
$V_{\text{In}}$ case with 85, 86 and 88~ps for the
$V_{\text{In}}$-$Si_{\text{In}}$, $V_{\text{In}}$-$O_{\text{N}}$ and
the $V_{\text{In}}$-3$O_{\text{N}}$, respectively. In the ratio
curves of the momentum densities shown in
Fig.~\ref{THEORY_Combined}(c) the peak maximum decreases with
increasing number of O ions, while the intensity in the spectral
range above 0.9~a.u. increases including the shoulder at 1.2~a.u.
and the peak at 3.4~a.u. The form of the
$V_{\text{In}}$-$O_{\text{N}}$ ratio curve resembles the case of
$V_{\text{In}}$ trapping with a reduced annihilation fraction of
$\eta \approx$~0.8. This can also be seen in the SW-plot
(Fig.~\ref{SW_THEORY}) where the characteristic points for all
calculated vacancy impurity complexes roughly fall on the
InN~-~$V_{\text{In}}$ line, meaning that their lineshape parameters
could be reproduced from the $V_{\text{In}}$ values assuming
incomplete annihilation fractions. The spectrum of the
$V_{\text{In}}$-$Si_{\text{In}}$ is very close the $V_{\text{In}}$
and hence hardly distinguishable in experiments. The case is
different for the ratio curve of the $V_{\text{In}}$-3$O_{\text{N}}$
which possesses distinct features with the shoulders at 1.2 and
3.6~a.u., respectively, which should be measurable in coincidence
Doppler measurements. The corresponding lineshape S (W) parameters
of the calculated vacancy-impurity complexes are 1.045 (0.83), 1.014
(0.92) and 1.049 (0.77) for the $V_{\text{In}}$-$O_{\text{N}}$,
$V_{\text{In}}$-3$O_{\text{N}}$ and
$V_{\text{In}}$-$Si_{\text{In}}$, respectively.

\subsection{Summary and discussion}
Our calculations can be summarized with following conclusions.
Isolated $V_{\text{N}}$ and pure $V_{\text{N}}$ complexes in the
neutral or positive charge state do not localize the positron
density and hence cannot be detected with positron annihilation
spectroscopy. When negatively charged, they might act as shallow
traps~\cite{Saarinen1998} for positrons. Isolated $V_{\text{In}}$
are efficient positron traps in InN with annihilation
characteristics (positron lifetime, ratio curve, lineshape
parameters) which are clearly distinguishable from the InN lattice
values. This is the case for all calculated defect complexes which
include at least one $V_{\text{In}}$. Apart from the
2$V_{\text{In}}$-$V_{\text{N}}$ and $V_{\text{In}}$-3$V_{\text{N}}$
complexes, their calculated positron lifetime values are very
similar with lifetime differences from 85~ps to 95~ps compared to
the InN lattice value. Due to the experimental accuracy their
distinction based on characteristic defect lifetimes is in practice
not feasible. The case is different for
2$V_{\text{In}}$-$V_{\text{N}}$ and $V_{\text{In}}$-3$V_{\text{N}}$
complexes, where lifetimes are higher with 109 and 99~ps difference
to the InN lattice,
respectively.\\
Distinction between the isolated $V_{\text{In}}$ and the calculated
complexes complexes can be achieved when considering their momentum
distributions. For the characteristic lineshape parameters of mixed
$V_{\text{In}}$-$nV_{\text{N}}$ complexes a clear deviation from the
InN~-~$V_{\text{In}}$ line is found in Fig.~\ref{SW_THEORY}, with a
strong increase of the S parameter with nearly constant W parameter
for an increasing number of $V_{\text{N}}$ around the
$V_{\text{In}}$. Its origin is revealed in the corresponding ratio
curves as a growing peak maximum and nearly constant area for
momenta between 1.5 and 3~a.u., where the dominant contributions to
the S respectively W parameter stem from. The most distinct feature
of these $V_{\text{In}}$-$nV_{\text{N}}$
complexes is identified in the disappearing shoulder at 1.2~a.u.\\
The characteristic points of the calculated vacancy-impurity and
pure $V_{\text{In}}$ complexes do not show such deviation but fall
on (an extension of) the InN~-~$V_{\text{In}}$ line. Accurate
determination of the annihilation fractions would be necessary for
an unambiguous identification of the 3$V_{\text{In}}$ complex. The
$V_{\text{In}}$-3$O_{\text{N}}$ complex exhibits some unique
features in its ratio curve, but $V_{\text{In}}$-$Si_{\text{In}}$
and 2$V_{\text{In}}$ are hardly distinguishable from the isolated
$V_{\text{In}}$.

\section{Experimental spectra and defect
identification}\label{Section_Experimental}
\begin{table}[b]
\caption{Studied samples and experimentally determined lineshape
parameters of the respective dominant positron trap. All parameters
are extrapolated to saturation trapping and divided by the reference
values measured in InN lattice.}
\begin{ruledtabular}
\begin{tabular}{cccccc}
 & & \textit{Layer: }& & \textit{Interface:} & \\[2pt]
ID & Sample &S/S$_{\text{ref}}$&W/W$_{\text{ref}}$&S/S$_{\text{ref}}$&W/W$_{\text{ref}}$\\[3pt]
\hline\\
I & MBE, irr. & 1.042 & 0.80 & 1.083 & 0.78 \\[3pt]
II & MBE, Si-doped & 1.051 & 0.83 & 1.077 & 0.81 \\[3pt]
III & MOCVD & 1.052 &  0.81 & /  & /\\[3pt]
\end{tabular}
\end{ruledtabular}
\label{table_SW_experimental}
\end{table}
\begin{figure}
\centering
\includegraphics[width=0.85\linewidth]{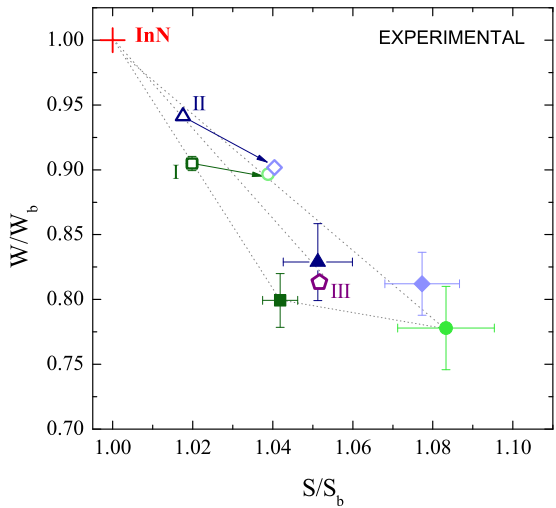}
  \caption{(Color online) Experimentally determined S and W values from conventional Doppler broadening setup for different
  samples (see Table~\ref{table_SW_experimental}). Open symbols correspond to directly measured values,
  closed symbols have been extrapolated using trapping
  fractions estimated from positron lifetime measurements. Light colors are used for points at high implantation energies, close to the interface (see arrow), while dark colors indicate points for low implantation energies, in the InN layer region.}
  \label{SW_EXPERIMENT}
\end{figure}
In order to identify the dominant positron traps in common InN
material we investigate a variety of as-grown and irradiated layers
that were grown by different growth methods. An overview of a
representative selection of measured samples is given in
Table~\ref{table_SW_experimental}. Sample I is MBE-grown
material~\cite{Tuomisto2007e} which has been irradiated with 2~MeV
He ions to a fluence of 8.9~$\times$~10$^{15}$cm$^{-2}$, sample II
and III are as-grown Si-doped~\cite{Schaff2004} and
undoped~\cite{Maleyre2004} InN layers deposited by MBE and MOCVD,
respectively. Details on the growth and characteristics of the
samples can be found
elsewhere~\cite{Tuomisto2007e,Reurings2010,Schaff2004,Rauch2010,Maleyre2004,Pelli2006}.
Sample I and II show a strong change in the Doppler broadening
signal when approaching the interface, which is a common feature in
several previously investigated InN
samples~\cite{Oila2004,Pelli2006,Rauch2010,Reurings2010}. Therefore,
the interface region is investigated separately in these two samples
(see section~\ref{Section_Interface}).\\
Conventional Doppler broadening (Fig.~\ref{SW_EXPERIMENT}) as well
as coincidence Doppler (Fig.~\ref{EXP_Combined}) spectra are
recorded for all samples. The recorded momentum distributions for
samples I and II are extrapolated to saturation trapping, using
annihilation fractions determined from previously recorded positron
lifetime measurements. An extensive description of the positron
lifetime data is published elsewhere~\cite{Rauch2010,Reurings2010}.
All experimental momentum distributions are shown as ratio curves
divided by the spectrum of a suitable reference for the InN lattice.
The reference sample has been carefully analyzed using positron
lifetime spectroscopy and no positron annihilation in trapped states
at vacancy defects is observed~\cite{Reurings2010a}. All samples are
measured perpendicular to the \textit{c}-axis.

\subsection{Irradiation-induced defects}\label{Section_Irradiated}
\begin{figure}
\centering
\includegraphics[width=1\linewidth]{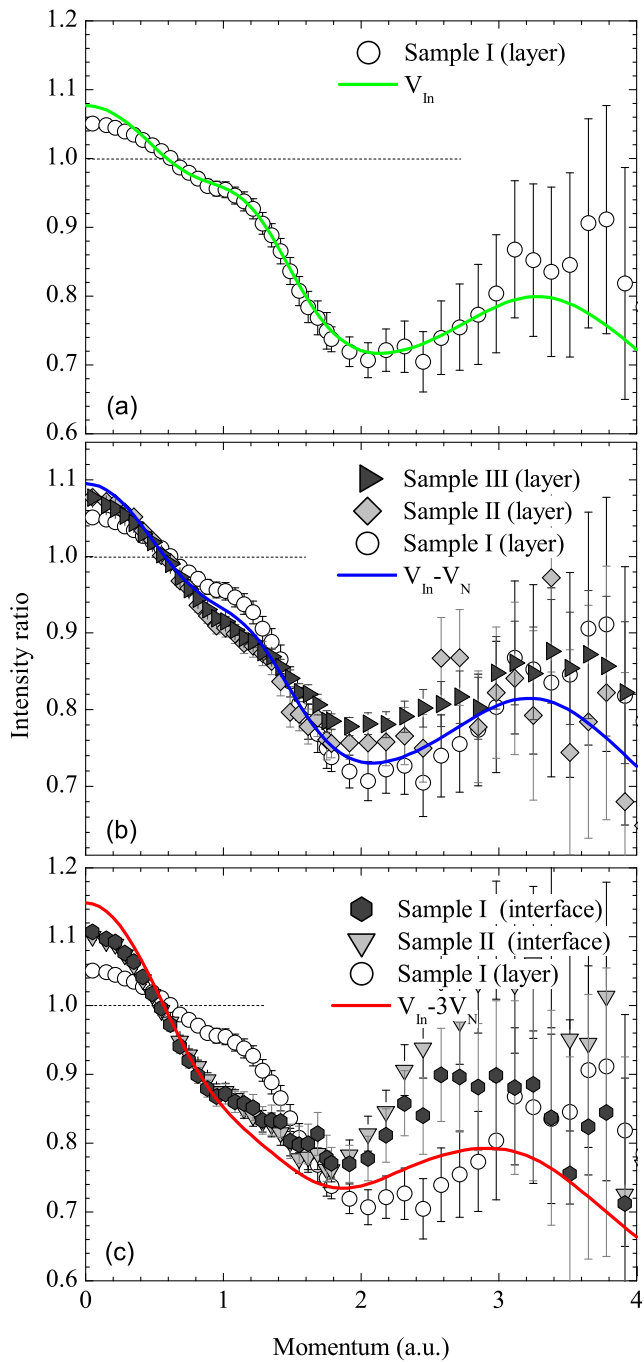}
  \caption{(Color online) Experimental coincidence Doppler spectra of the investigated samples in the layer (a,b) and interface (c) region. The data has been divided by a suitable reference spectrum for the InN lattice.
Computational ratio curves are shown for comparison.}
  \label{EXP_Combined}
\end{figure}

Positron lifetime measurements~\cite{Reurings2010} of the layer
region of sample I show one dominant positron trap with a
characteristic lifetime of 260~ps, and an annihilation fraction of
$\eta=0.47$ is determined for the trap. The recorded lineshape
parameters from conventional Doppler broadening measurements are
extrapolated accordingly and displayed in Fig.~\ref{SW_EXPERIMENT}.
Characteristic line shape parameters of S~=~1.042 and W~=~0.80 are
evaluated for the dominant defect in this region.\\
The extrapolated ratio curve of sample I is shown in
Fig.~\ref{EXP_Combined}(a). When comparing the experimentally
determined defect ratio curve to the calculated momentum
distributions in section~\ref{computational} we find good agreement
with the spectrum of the isolated $V_{\text{In}}$ for the most of
the spectral range. In the central region of the peak slightly
higher intensities are found in the calculated spectrum compared to
the experimental one. This region is mostly sensitive to the size of
the open volume of the positron trap, with higher intensities for
larger volumes. The calculated 2$V_{\text{In}}$ and
$V_{\text{In}}$-$O_{\text{N}}$ complexes possess an overall rather
similar shape of the ratio curve and based on this also have to be
considered as possible sources of the defect signal. Nevertheless,
the available lifetime data does not support the reduced
annihilation fraction which would be required for an identification
of the experimental spectrum with the
$V_{\text{In}}$-$O_{\text{N}}$. Although the differences between the
$V_{\text{In}}$ and 2$V_{\text{In}}$ complex are more subtle and
hence the 2$V_{\text{In}}$ cannot be ruled out contributing to the
signal, the ratio curve of the $V_{\text{In}}$ gives the best
overall approximation of the experimental spectrum. Therefore we
identify the positron trap created in irradiated InN with the
isolated $V_{\text{In}}$.

\subsection{Defects in as-grown samples}\label{Section_AsGrown}
The experimentally determined S and W parameters from conventional
Doppler broadening for the as-grown samples II and III are displayed
in Fig.~\ref{SW_EXPERIMENT}. Positron lifetime
measurements~\cite{Rauch2010} show a annihilation fraction of
$\eta=0.34$ in the dominant positron trap for sample II. Based on
this, the characteristic lineshape parameters for this defect are
estimated as 1.051 and 0.83 for S and W, respectively. This is close
to the as-measured line shape parameters of sample III, i.e.
S~=~1.052 and W~=~0.81. For this sample no lifetime data is
available. In Fig.~\ref{EXP_Combined}(b) the respective ratio curves
recorded with coincidence Doppler measurements are displayed. The
extrapolated ratio curve of both sample II and the as-measured
spectrum of sample III show a very similar line shape. The bigger
scatter in the former is due to a smaller annihilation
fraction.\\
Compared to sample I, the as-grown samples II and III show several
differences in their ratio curves. First, the intensity in the peak
center region is clearly increased. The intensity difference to the
InN lattice is thereby magnified by about 35~\% compared to the
spectrum of sample I. This is supported by very accurate statistics
in this spectral region. Second, a significant decrease of the
shoulder at 1.2~a.u. is visible, also with high statistical
accuracy. Third, the drop at 2~a.u. is less pronounced, followed by
slightly higher intensities in the high momentum region of the
spectrum. Nevertheless, stronger scatter starts to dominate
this region.\\
A comparison with the calculated defect spectra in
section~\ref{Section_Theory} reveals that these changes coincide
with the effects of the decoration of a $V_{\text{In}}$ by
$V_{\text{N}}$, as presented in Fig.~\ref{THEORY_Combined}(b).
Especially the characteristic decrease of the shoulder at 1.2~a.u.
in ratio curves of the experimental spectra cannot be correlated
with any other calculated vacancy defect complex (see additionally
Hautakangas \textit{et al.}~\cite{Hautakangas2006a}). This is also
expressed in the observed deviation of the characteristic lineshape
parameters of sample II and III from the line determined by the
characteristic points of sample I and the InN lattice in
Fig.~\ref{SW_EXPERIMENT}. A similar trend can be observed for the
calculated $V_{\text{In}}$-$nV_{\text{N}}$ parameters in
Fig.~\ref{SW_THEORY}. Judging from the amount of the observed
changes in the ratio plots an identification of the experimental
spectra with $V_{\text{In}}$-$V_{\text{N}}$ is most feasible, with
possible influence from the $V_{\text{In}}$-2$V_{\text{N}}$. This
assignment is in good agreement with the positron lifetime
data~\cite{Rauch2010}.

\subsection{Defects at the Interface}\label{Section_Interface}
A strong change in the Doppler broadening signal is observed for
sample I and II close to the interface region, as visible in
Fig.~\ref{SW_EXPERIMENT}. From positron lifetime data in this
region~\cite{Reurings2010,Rauch2010} we determine a positron
annihilation fraction of $\eta=0.47$ and $\eta=0.52$ at the dominant
positron traps in samples I and II, respectively. The characteristic
defect lifetime at the interface is slightly increased compared to
the layer region. The extrapolated line shape parameters are
determined as 1.083 (0.78) and 1.077 (0.81) for S (W) in samples I
and II, respectively, and show an even stronger deviation from the
line defined by the characteristic points of sample I and the InN
reference as already observed in section~\ref{Section_AsGrown}. The
characteristic SW points of both interfaces coincide within the
statistical accuracy. This indicates that the same dominant positron
trap is present in both samples, which is also shown in the
extrapolated ratio curves shown in Fig.~\ref{EXP_Combined}(c). In
both samples a strong increase in the peak center intensity to about
1.12 is visible, which is over 2-fold compared to that observed in
the irradiated layer. Additionally, the signal drops straight to the
minimum at 2~a.u. without showing anymore the shoulder which is
visible in the layer region of both samples. The observed trends are
qualitatively very similar to the ones described in the previous
section for the layer region of samples II and III, but intensified.
Therefore, we identify the induced changes with an increase in the
decoration of $V_{\text{In}}$ with $V_{\text{N}}$. When comparing to
the calculated momentum distributions in
section~\ref{Section_Theory}, best agreement is found for the
spectrum of the $V_{\text{In}}$-3$V_{\text{N}}$ complex. Results
from both conventional Doppler broadening (Fig.~\ref{SW_EXPERIMENT})
and positron lifetime spectroscopy additionally support this
assignment~\cite{Rauch2010,Reurings2010a}.

\subsection{Summary and discussion}\label{Section_Discussion}
Based on the above presented data we are able to identify the
dominant positron trap created in high-energy particle irradiation
of MBE grown InN layers as the isolated $V_{\text{In}}$, while in
as-grown MBE and MOCVD material the observed defect is a mixed
$V_{\text{In}}$-$V_{\text{N}}$ vacancy complex. The changes at the
interface in both irradiated material and as-grown layers are
assigned to the formation of larger $V_{\text{In}}$-$nV_{\text{N}}$
complexes with an average of about 3 $V_{\text{N}}$ surrounding the
$V_{\text{In}}$. It has to be noted that the positron implantation
profile for the chosen interface point (12~keV) is already
considerably broadened and hence signal from a wide area of the
layer is averaged. The observed complexing of $V_{\text{In}}$ and
$V_{\text{N}}$ is in good agreement~\cite{Rauch2011b} with existing
data on the electron mobility in these samples~\cite{Jones2007} and
might be the reason behind observed superior transport
properties.\\
The measured Doppler broadening of the annihilation $\gamma$
radiation is caused by the momentum component of annihilating e-p
pairs along the detection direction~\cite{Saarinen1998}. All our
experimental spectra are recorded perpendicular to the
\textit{c}-axis but the exact detection direction in the c-plane is
not identified. We compare the experimental spectra to computational
momentum distributions integrated over the $m$-plane. The induced
uncertainty could in principle cause difficulties for the exact
defect identification, as small differences between the different
lattice directions are present~\cite{Rauch2011}. Nevertheless, these
are minor compared to differences between the calculated spectra of
different defect complexes and hence pose no problem in our case.
This counts especially when interpreting changes in
the momentum distributions rather than absolute values.\\
Our data accentuates the advantage of high-quality coincidence
Doppler~\cite{Lynn1977} spectra for the identification of defect
identities. Apart from the advanced experimental resolution compared
to conventional Doppler measurements which helps to distinguish
defect-induced changes (see Fig.~\ref{THEORY_Combined}(a)),
important features in the spectra might be lost when only regarding
the integrated lineshape parameters S and W. This is the case, e.g.,
for the observed disappearance of the shoulder at 1.2~a.u. which is
identified as an unique feature of the
$V_{\text{In}}$-$nV_{\text{N}}$ complexes. The commonly used
integration windows for the S and W parameters cannot include this
momentum area, in order to avoid correlation
effects~\cite{Saarinen1998}. Irrespective of this, a shift of the
lower limit of the W-parameter window to 2.2~a.u. (for a detector
resolution of 1.24~keV) would be beneficial in future positron
experiments in InN in order to maximize the visibility of
defect-induced changes (see Fig.~\ref{THEORY_Combined}).

\section{Conclusion}
We present the identification of dominant vacancy-type positron
traps in different representative InN samples by combining positron
lifetime and Doppler broadening spectroscopy with ab-initio
calculations of the positron annihilation characteristics.
Calculated momentum distributions of annihilating electron positron
(e-p) pairs in different vacancy type defects are compared to high
resolution measurements of the Doppler broadened e-p annihilation
radiation acquired with the coincidence Doppler technique. We find
that isolated $V_{\text{N}}$ or pure $V_{\text{N}}$ complexes do not
trap positrons. During high-energy particle irradiation isolated
$V_{\text{In}}$ are created, which are the dominant positron trap in
such samples at room temperature. In as-grown InN samples on the
other hand, $V_{\text{In}}$ do not occur isolated but complexed with
one or more $V_{\text{N}}$. The observed changes of the Doppler
broadening signal close to the interface are identified with an
increasing amount of $V_{\text{N}}$ surrounding the $V_{\text{In}}$.
The characteristic lifetimes, ratio curves and S and W parameters
for the identified defects are determined.

\begin{acknowledgements}
The authors thank the groups of W. Schaff at Cornell University, J.
Speck at University of California Santa Barbara, and O. Briot at
Universit\'{e} Montpellier II for the generous supply of InN
samples. F. Reurings is acknowledged for his help with positron
lifetime measurements. This work has been supported by the European
Commission under the 7th Framework Program through the Marie Curie
Initial Training Network RAINBOW, Contract No. PITN-Ga-2008-213238.
\end{acknowledgements}

%\bibliography{Positron}
%\bibliographystyle{apsrev}

\end{document}